\documentclass[usenatbib,useAMS,letters]{mnras}

\usepackage{natbib}
\usepackage{amsmath}
\usepackage{graphicx}
\usepackage{color}
\usepackage{mathrsfs}
\usepackage{epsfig}
\usepackage{comment}
\usepackage{ulem}

\usepackage{graphicx}
\usepackage[hang,small,bf]{caption}
\usepackage[subrefformat=parens]{subcaption}
\usepackage{amssymb}
\usepackage[T1]{fontenc}

\newcommand{\msun}{{M}_{\odot}}
\newcommand{\BHNS}{{BH1st--NS2nd}}
\newcommand{\NSBH}{{NS1st--BH2nd}}

\newcommand{\rsun}{{\rm R}_\odot}
\newcommand{\zsun}{Z_\odot}

\newcommand{\beq}{\begin{equation}}
\newcommand{\eeq}{\end{equation}}

\defcitealias{K14}{K14}
\defcitealias{K16}{K16}
\defcitealias{Visbal_2015}{VHB15}

\usepackage{url} 

\title[Constraints on Pop I/II NS-BH formation]
{Constraints on Population I/II neutron star–black hole binary formation by gravitational wave and radio observations}

\author[T. Kinugawa et al.]
{Tomoya Kinugawa$^{(1)}$\thanks{E-mail: kinugawa@icrr.u-tokyo.ac.jp},  Takashi Nakamura$^{(2)}$, and Hiroyuki Nakano$^{(3)}$\\
\\
$^{1}$Institute for Cosmic Ray Research, The University of
  Tokyo, Kashiwa, Chiba 277-8582, Japan\\
$^{2}$Department of Physics, Graduate School of Science, Kyoto University,
Kyoto 606-8502, Japan\\
$^{3}$Faculty of Law, Ryukoku University, Kyoto 612-8577, Japan}

\begin{document}

\date{\today}
\maketitle
\begin{abstract}
Two neutron star (NS)–black hole (BH) binaries,
GW200105 and GW200115 found 
in the LIGO/Virgo O3b run have smaller BH mass
of 6--9\,$\msun$ which is consistent with Population I and II origin.
Our population synthesis simulations using $10^6$ Population I and II binaries with appropriate initial parameters show 
consistent binary mass, event rate, and no detection of radio pulsar (PSR) and BH binaries in our galaxy so far. Especially, we found possible progenitors of GW200105 and GW200115 which were formed at redshift $z=0.15$ and $z=1.6$ with binary mass of $(34\msun,\, 9.2\msun)$ and $(23.7\msun,\, 10.6\msun)$, respectively. The final masses of these binaries are $(6.85\msun,\,2.14\msun)$ and $(6.04\msun,\,1.31\msun)$ which look like $(9.0_{-1.7}^{+1.7}\msun,\, 1.91_{-0.24}^{+0.33}\msun)$ of GW200105 and $(5.9_{-2.5}^{+2.0}\msun,\,1.44_{-0.29}^{+0.85}\msun)$ of GW200115, respectively.
We also estimate that 2.68-19.7 PSR-BH binaries in our galaxy will be observed by SKA. The existence of NS-BHs in our galaxy can be 
confirmed in future SKA era. Using the GW observation of NS-BH mergers and the radio observation of PSR-BHs in future, we can get more severe constraints on the NS-BH formation process.
\end{abstract}


\section{Introduction}

Two gravitational wave (GW) events
of neutron star–black hole (NS-BH) coalescences,
GW200105\_162426 (abbreviated as GW200105)
and GW200115\_042309 (abbreviated as GW200115)
were observed by the LIGO–Virgo detector network~\citep{LIGOScientific:2021qlt}.
Table~\ref{tab:events} is the summary of the events.
Here, we present only the chirp mass
$M_{\rm chirp} = (m_1 m_2)^{3/5}/M^{1/5}$
(where $M=m_1+m_2$), 
primary mass $m_1$ and secondary mass $m_2$ 
although the other parameters (spin etc.)
have been estimated
(e.g., an effective spin parameter
defined with nondimensional spin parameters
($\chi_{1,z}$ and $\chi_{2,z}$)
parallel to the orbital angular momentum, 
$\chi_{\rm eff}= (m_1/M)\chi_{1,z}+(m_2/M)\chi_{2,z}
= 0.00^{+0.13}_{-0.18}$ for GW200105
and 
$-0.15^{+0.24}_{-0.42}$ for GW200115~\citep{2021arXiv211103606T},
see also \cite{2021ApJ...922L..14M} for the BH spin 
of GW200115)~\footnote{To extract more detailed
information from these binaries, we need multiband 
GW observations, i.e., ground-based and space-based detectors
(see e.g.,~\cite{2021arXiv210808490L}
as an extension 
of~\cite{2018PTEP.2018g3E01I,2021Univ....7...53N})}.
The NS-BH merger rate density
(combined with analyses based on the two events,
and including less significant search triggers)
was estimated as $12$--$242\,{\rm yr}^{-1}{\rm Gpc}^{-3}$.
This estimation has been updated
to $7.4$--$320\,{\rm yr}^{-1}{\rm Gpc}^{-3}$
in~\cite{2021arXiv211103634T} 
using the third Gravitational-wave Transient Catalog
(GWTC-3)~\citep{2021arXiv211103606T}
(see also GWTC-2.1~\cite{2021arXiv210801045T}
and the fourth Open Gravitational-wave Catalog 
(4-OGC)~\citep{2021arXiv211206878N}).

After the announcement of~\cite{LIGOScientific:2021qlt}, 
various works on the population of these binaries 
have appeared.
The works includes studies on 
primordial BH (PBH) scenarios~\citep{2021arXiv210700450W,2021arXiv210811740C,2021arXiv211009509S},
isolated binary evolution scenarios~\citep{2021ApJ...920...81S,2021ApJ...920L..13B,2021ApJ...918L..38F}
(see also~\cite{2021ApJ...920L..20Z}),
the mass distribution of NSs/BHs in GW events~\citep{2021ApJ...921L..25L,2021arXiv210714239M},
hierarchical population inferences~\citep{2021ApJ...923...97L},
the tilt angle of the BH spin~\citep{2021arXiv210810184G}, 
prediction of short-duration gamma-ray bursts~\citep{2021arXiv210909714M},
highly unequal mass components~\citep{2021arXiv211001393A},
quadruple-star systems~\citep{2021arXiv211014680V},
the lower mass gap between NSs and BHs~\citep{2021arXiv211103498F},
dynamical interactions in low-mass young star clusters~\citep{2021arXiv211106388T},
and prediction on GW event rates~\citep{2021arXiv211113704W}
observed by a space-based detector, LISA~\citep{2017arXiv170200786A}
(see also, \cite{2021arXiv210810885B,2021arXiv211205763B}).
\cite{2021arXiv210714239M} gave us a great summary
on merger rates of compact object binaries, i.e.,
NS-NSs, BH-BHs and NS-BHs
for GW observations and various formation channels.
Using information from the NS-BH GW events 
and radio pulsar (PSR) surveys in which 
no PSR-BH has been observed,
\cite{2021arXiv210904512P} derived 
an upper limit (95\% CL) of $\sim 150$ PSR-BHs
in our galaxy with the beaming direction to the Earth.
\cite{2021MNRAS.504.3682C}
calculated that $1$--$80$ PSR-BHs are detectable by SKA observation.

\begin{table}
\caption{Abbreviated event name, chirp mass $M_{\rm chirp}$, primary mass $m_1$ and secondary mass $m_2$ in unit of the solar mass, $M_{\odot}$ are from GWTC-3~\citep{2021arXiv211103606T}. Each value is shown with the 90\% credible interval.}
\label{tab:events}
\begin{center}
\begin{tabular}{cccc}
\hline
Event name & $M_{\rm chirp}$ & $m_1$ & $m_2$ \\
\hline
GW200105 & $3.42_{-0.08}^{+0.08}$ 
& $9.0_{-1.7}^{+1.7}$ & $1.91_{-0.24}^{+0.33}$ \\
GW200115 & $2.43_{-0.07}^{+0.05}$
& $5.9_{-2.5}^{+2.0}$ & $1.44_{-0.29}^{+0.85}$ \\
\hline
\end{tabular}
\end{center}
\end{table}

In~\cite{Kinugawa2017} (hereafter Paper I),
we discussed the merger rate of Population (Pop) I, II and Pop III NS-BH binaries
by using population synthesis Monte Carlo simulations,
including the kick of NSs~\citep{1994Natur.369..127L,1997MNRAS.291..569H}
(see also \cite{2005MNRAS.360..974H,2017A&A...608A..57V}).
From Table 3 of Paper I, we found that the merger rates of Pop I, II and Pop III are $6.38$--$19.7\,{\rm yr}^{-1}{\rm Gpc}^{-3}$, and $0.956$--$1.25\,{\rm yr}^{-1}{\rm Gpc}^{-3}$, respectively.
Therefore, the merger rate of Pop I and II NS-BHs in Paper I is consistent with the LIGO-Virgo result of $7.4$--$320\,{\rm yr}^{-1}{\rm Gpc}^{-3}$
in~\cite{2021arXiv211103634T}.
This implies that there are many PSR-BHs made by Pop I and II binary systems.
In this Letter, based on the analysis in Paper I,
{we also calculate the Pop I, II NS-BHs merger rate with other parameters} and estimate the number of PSR-BHs detected by current and future observations.

\section{Analysis}

Theoretically NS-BH binaries can be formed in a certain evolution of binary stars. BH is usually formed first
and next is NS
because BH is usually formed from more massive star with shorter evolution time than the progenitor of NS. This means that 
some of them can be observed as a binary radio PSR with BH. However, no such object has been observed so far. 
This situation of no observed radio counter object of NS-BH binary 
is allowed if the expected number of NS-BH radio PSRs observed by existing radio telescopes is smaller than
or compatible with the order of unity. 


\subsection{Population synthesis method}
In Paper I, we performed population synthesis Monte Carlo simulations of Pop I, II and III binary stars using $10^6$ binaries calculated by a modified BSE code \citep{Hurley_2002,Kinugawa2014,Kinugawa:2015nla} for each given metallicity of $Z=\zsun,\, 10^{-0.5}\zsun,\, 10^{-1}\zsun,\, 10^{-1.5}\zsun,\, 10^{-2}\zsun$ and $Z=0$ (i.e., Pop III stars) where $\zsun$ is the metallicity of the Sun.
{We used the Madau star-formation rate (SFR) \citep{Madau2014}, and calculated the metallicity evolution using the galaxy mass-metallicity relation evaluated by simulation \citep{Ma2016} and the galaxy mass distribution fitted by the Shechter function \citep{Fontana2006}.
We assumed the metallicity $Z$ is changed as a function of redshift $z$ taking the intermediate value in log scale such as
$Z=10^{-1.75}\zsun~(z=6.745)$, $10^{-1.25}\zsun~(z=5.168)$, $10^{-0.75}\zsun~(z=2.528)$, and $10^{-0.25}\zsun~(z=0.096)$.
We choose the binary initial conditions, i.e., the primary mass $M_1$, the
mass ratio $q=M_2/M_1$, the separation $a$, and the eccentricity $e$ when the binary is born. We adopt the Salpeter initial mass function (IMF) ($5\msun<M_1<140\msun$), the flat initial mass ratio function ($0.1\msun/M_1<q<1$),
the logflat initial separation function ($a_{\rm min}<a<10^6\rsun$), and the thermal initial eccentricity distribution function ($0<e<1$) for Pop I, II binaries.
On the other hand, we adopt the flat IMF and the different minimum mass ($10\msun$) for Pop III stars.}
We adopt in the simulation that when NS is formed it gets kick velocity of Maxwellian shaped absolute value with random direction. We adopt  two values of velocity dispersion of the kick velocity $\sigma_k$ of $265$\,km/s and $500$\,km/s. The former value of the kick velocity is that of the observed single PSRs \citep{2005MNRAS.360..974H} and the latter is that of young PSRs~\citep{2017A&A...608A..57V}.

\begin{table}
\caption{Parameters for eight models.}
\label{tab:model}
\begin{center}
\begin{tabular}{ccccc}
\hline
 model &  $\sigma_k$ & $\alpha\lambda$&$\beta$& BH kick\\
 \hline
  Standard & 265\,km/s & 1 & 0 & off\\
  $\sigma_k=500$\,km/s & 500\,km/s & 1 & 0 & off\\
  $\alpha\lambda=0.1$ & 265\,km/s & 0.1 & 0 & off\\
  $\alpha\lambda=0.5$ & 265\,km/s & 0.5 & 0 & off\\
  $\alpha\lambda=10$ & 265\,km/s & 10 & 0 & off\\
  $\beta=0.1$ & 265\,km/s &1 & 0.1 & off\\
  $\beta=0.5$ & 265\,km/s & 1 & 0.5  & off\\
  BH kick & 265\,km/s & 1 & 0 & on\\
 \hline
\end{tabular}
\end{center}
\end{table}

\begin{table*}
\caption{Pop I, II NS-BH merger rate at the present day for each model.}
\label{tab.NS-BH merger}
\begin{center}
\begin{tabular}{ccccccccc}
\hline
  & Standard & $\sigma_k=500$\,km/s & $\alpha\lambda=0.1$ & $\alpha\lambda=0.5$& $\alpha\lambda=10$ & $\beta=0.1$& $\beta=0.5$ & BH kick \\
 \hline
  NS-BH merger rate [\,$\rm Gpc^{-3}~yr^{-1}$\,] &  19.7 & 6.38 & 19.1 & 23.2 & 1.07 &18.0& 19.0 & 5.78\\
 \hline
\end{tabular}
\end{center}
\end{table*}

In this Letter, we also calculate the Pop I, II NS-BH evolution with other binary parameters, i.e., the common envelope parameters $\alpha\lambda$, and the mass loss fraction $\beta$ during the Roche lobe over flow  \citep{Hurley_2002,Kinugawa2014, Kinugawa:2015nla}. 
{In Paper I, we assumed $\alpha\lambda=1$ and $\beta=0$ and no natal kick for BH formations.
Here, we also consider the cases of $\alpha\lambda=0.1,~0.5,~10$, $\beta=0.1,~0.5$ and the BH natal kick. Then, each model is characterized by the values of $\sigma_k$,
$\alpha\lambda$, $\beta$, and the switch of the BH natal kick. For simplicity,
we treat only eight models defined by the values of $\sigma_k$,
$\alpha\lambda$, $\beta$, and the switch of the BH natal kick (see Table~\ref{tab:model}). 
We consider 1) two models with different values of $\sigma_k =265$\,km/s or 500\,km/s, but the same values of $\alpha\lambda=1$ and $\beta=0$ with no BH natal kick, 2) three models with different values of $\alpha\lambda=0.1$, $0.5$ and $10$, but the same values of $\sigma_k=$265\,km/s and $\beta=0$ with no BH natal kick, 3) two models with $\beta=0.1$ and 0.5, but $\alpha\lambda=1$ and $\sigma_k=265$\,km/s with no BH natal kick,
and 4) the BH natal kick model with $\sigma_k =265$\,km/s, $\alpha\lambda=1$ and $\beta=0$.}
\
\begin{figure}
  \begin{center}
    \includegraphics[width=\hsize]{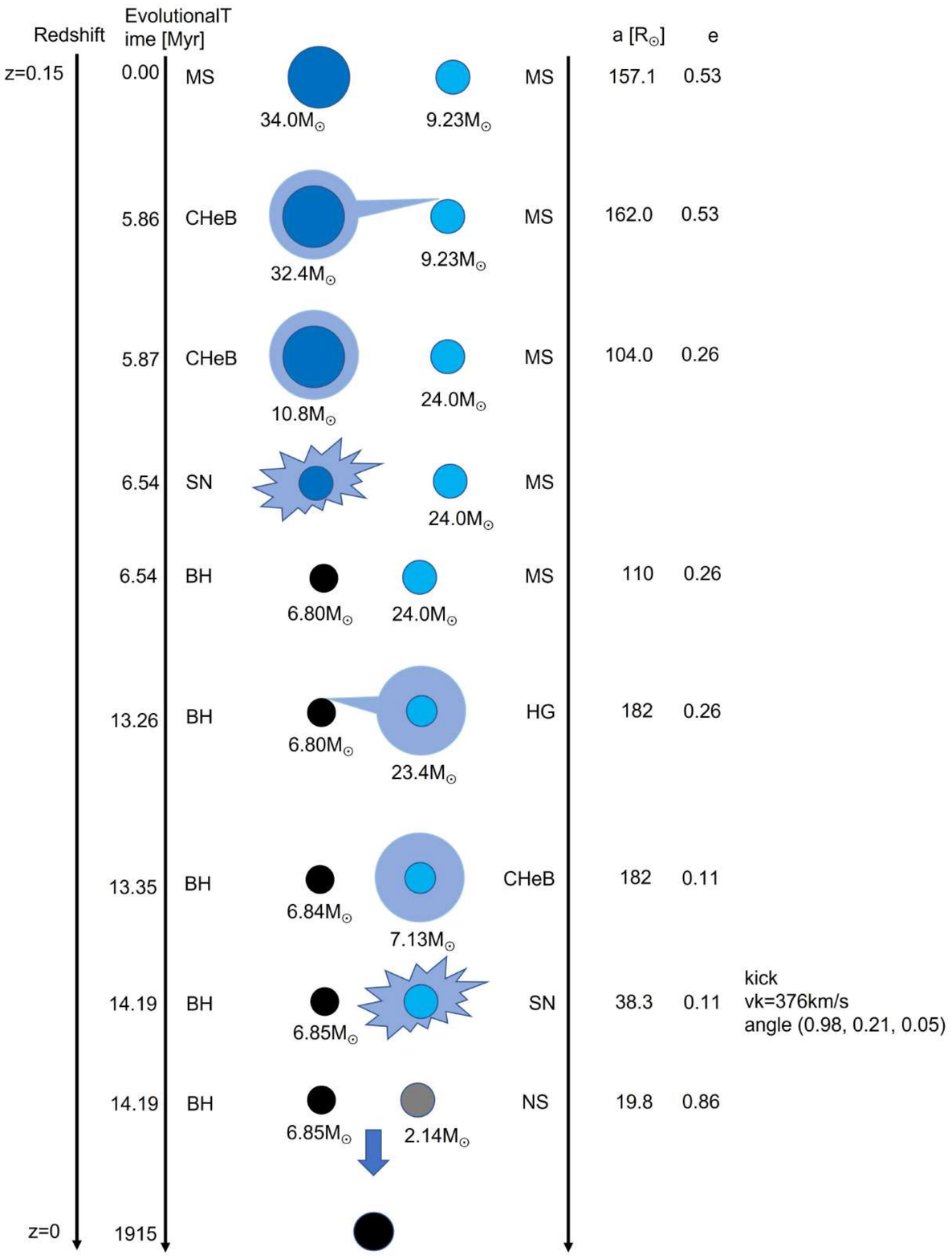}
  \end{center}
  \caption{The formation of a NS-BH with $m_1=6.85\,\msun$ and $m_2=2.14\,\msun$ which looks like GW200105 shown in Table~\ref{tab:events}.
  The pulsar kick velocity is $376$\,km/s with kick angle $(0.98,\, 0.21,\, 0.05)$ where the kick angle is defined in Figure~A1 of~\citet{Hurley_2002}. MS, CHeB SN, BH, HG, and NS means a main sequence phase, a core Helium burning phase, a supernova, a black hole, a Hertzsprung gap phase, and a neutron star, respectively.}
  \label{fig:GW200105}
\end{figure}
\begin{figure}
  \begin{center}
    \includegraphics[width=\hsize]{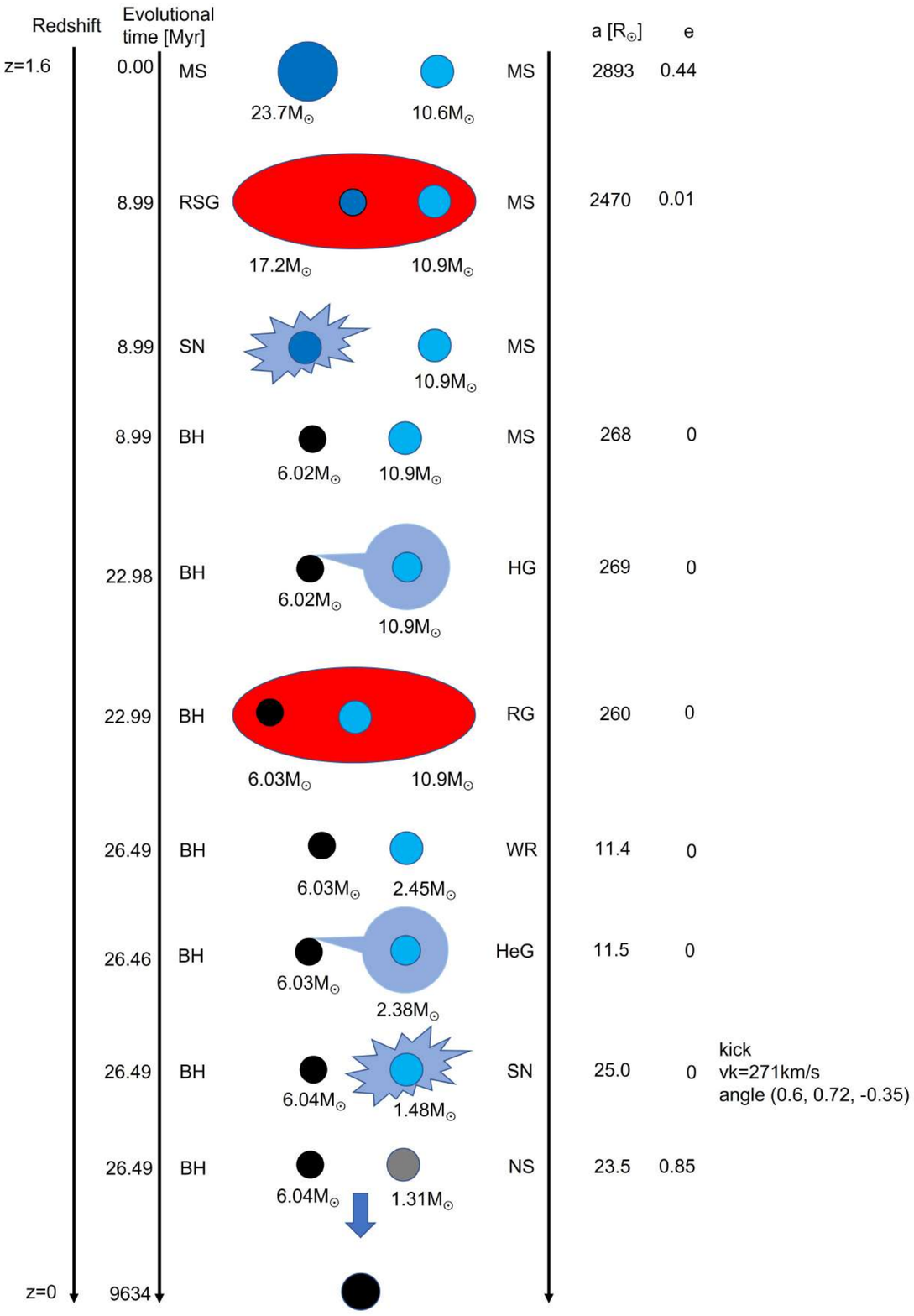}
  \end{center}
  \caption{The formation of a NS-BH~with $m_1=6.04\,\msun$ and $m_2=1.31\,\msun$ which looks like GW200115 shown in Table~\ref{tab:events}. The pulsar kick velocity is $271$\,km/s with kick angle $(0.6,\, 0.72,\, -0.35)$ where the kick angle is defined in Figure~A1 of~\citet{Hurley_2002}. RSG, RG, WR, and HeG mean a red super giant, a red giant, a Wolf-Rayet star, and a Helium giant, respectively.}
  \label{fig:GW200115}
\end{figure}
Table~\ref{tab.NS-BH merger} shows the merger rates of NS-BHs at the present day for each model.
{We show two examples from our simulation for $\sigma_k =265$\,km/s, $\alpha\lambda=1$ and $\beta=0$ with no BH natal kick}: the first one shown in Figure~\ref{fig:GW200105}
is a NS-BH binary formed at redshift $z=0.15$ with $m_1=6.85\,\msun$ and $m_2=2.14\,\msun$ which looks like GW200105, 
while the second one shown in Figure~\ref{fig:GW200115} is that of a NS-BH binary formed at $z=1.6$ with $m_1=6.04\,\msun$ and $m_2=1.31\,\msun$ which looks like GW200115.
These examples are made from binaries with $Z=10^{-0.5}\zsun$  which were born within a range of $z=0.1$--$2.5$. Note here that binaries with $Z=10^{-0.5}\zsun$ dominate the merger rate of NS-BHs at $z=0$ as shown in Figures 2 and 3 of Paper I.

\subsection{detectability of Pulsar- black hole binaries}
\begin{figure}
\includegraphics[width=\hsize]{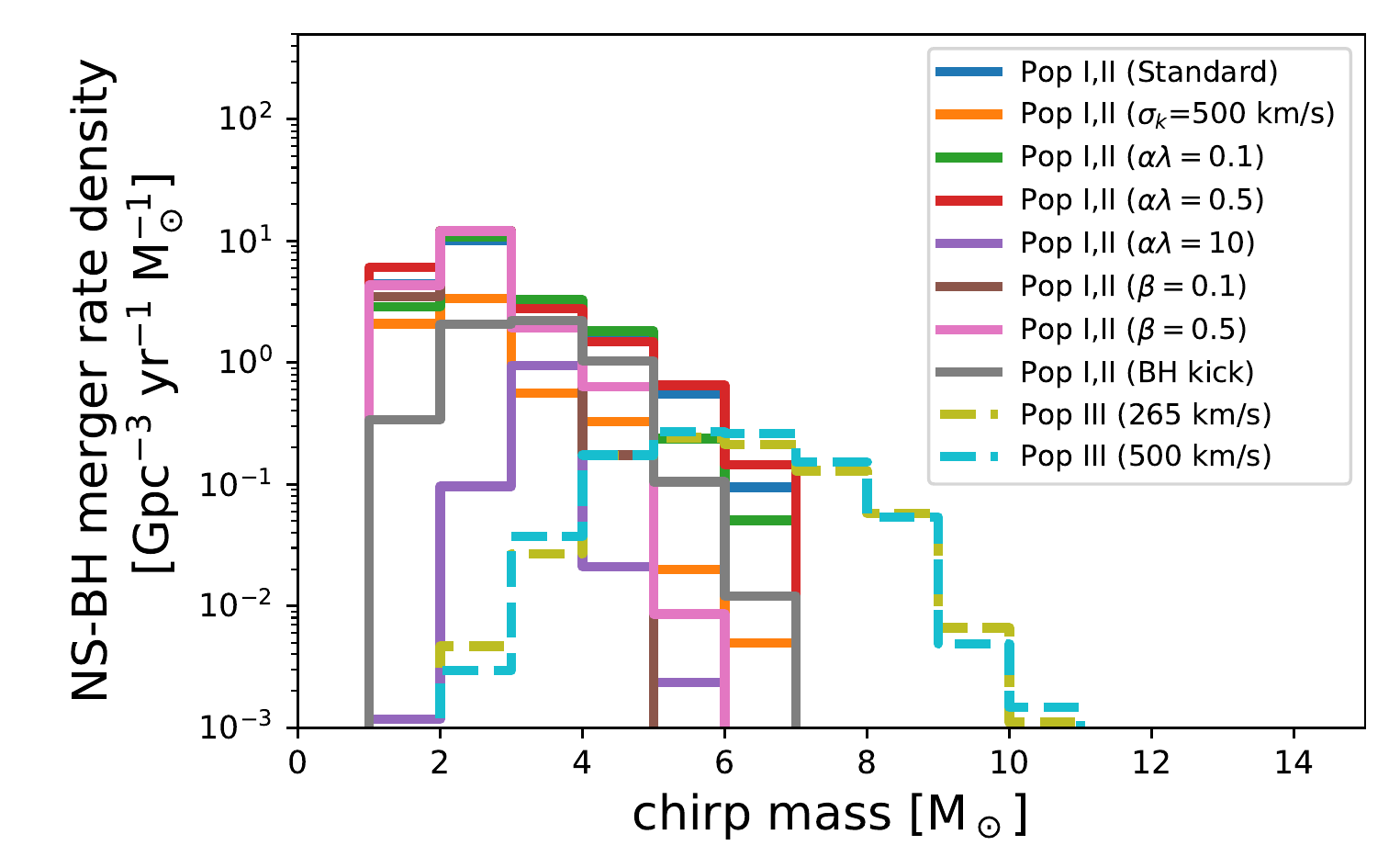}
    \caption{{Chirp mass distributions of detectable NS-BHs at $z=0$. The solid lines show the cases of Pop I, II NS-BH merger rates for each model, respectively. The dashed lines show the merger rates of Pop III NS-BHs from Paper I, respectively.}}
\label{fig:chirp}
\end{figure}

Table~\ref{tab:M_dep} of this Letter shows metallicity dependence on numbers of NS-BH~formation in $10^6$ Monte Carlo simulations. The numbers in parenthesis are NS-BHs which merge within the Hubble time. We treat two velocity dispersion models of the kick velocity of $\sigma_k =265$\,km/s, and $\sigma_k=500$\,km/s. 
Figure~\ref{fig:chirp} shows the NS-BH merger rate densities for each model. The peak of chirp mass distribution of detectable NS-BHs is around 2--3\,$\msun$, which is consistent with the observation of GW200105 and GW200115 since their chirp mass are $3.42\,\msun$ and $2.43\,\msun$, respectively.

\begin{table*}
\caption{Metallicity dependence on numbers of NS-BH formation in $10^6$ Monte Carlo simulations. The numbers in parentheses are NS-BHs which merge within the Hubble time. We treat five models with kick velocity of $\sigma_k =265$\,km/s and $\sigma_k=500$\,km/s, $\alpha\lambda=0.1$ and $10$,
and $\beta=0.5$.}
\label{tab:M_dep}
\begin{center}
\begin{tabular}{cccccc}
\hline
 $Z$ & $\zsun$ & $10^{-0.5}\zsun$&$10^{-1}\zsun$&$10^{-1.5}\zsun$&$10^{-2}\zsun$\\
 \hline
  NS-BH [\,Standard\,] & 147 (15) & 598 (191) & 1295 (524) & 1686 (755) & 1896 (862) \\
  NS-BH [\,$\sigma_k=500$\,km/s\,] & 32 (2) & 169 (67) & 416  (213) & 576 (377) & 617 (401) \\
  NS-BH [\,$\alpha\lambda=0.1$\,] & 128 (48) &362 (148) & 1312 (518) & 1941 (845) & 1932 (840) \\
  NS-BH [[\,$\alpha\lambda=0.5$\,] & 203 (29) & 651 (283) & 1428 (665) & 1724 (790) & 1907 (755) \\
  NS-BH [\,$\alpha\lambda=10$\,] & 20 (0) & 119 (14) & 383 (22) & 928 (100) & 1271 (200) \\
   NS-BH [\,$\beta=0.1$\,] & 60 (9) & 637 (191) & 1443 (556) & 1938 (704) & 2242 (849) \\ 
  NS-BH [\,$\beta=0.5$\,] & 65 (5) & 627 (190) & 1404 (547) & 1828 (720) & 2060 (842) \\
  NS-BH [ BH kick ] & 44 (16) & 124 (52) & 437 (262) & 782 (482) & 935 (503) \\
 \hline
\end{tabular}
\end{center}
\end{table*}

Figures \ref{fig:GW200105} and \ref{fig:GW200115} suggest that PSR-BH binaries should be observed in a certain time of the universe for each NS-BH binary formed in our scenario. The typical maximum age of observable radio PSRs is $\sim 5\times 10^7$\,yrs from the PSR death line 
for the magnetic field strength of $\sim 10^{12}$\,G which is typical for newborn PSRs and PSRs in high mass X-ray binaries (HMXBs)~\citep{Enoto2019}. 
We first estimate the number of NS-BHs formed
from $5\times 10^7$\,yrs ago up to now in our galaxy.
Using the first column of Table~\ref{tab:M_dep} for $Z=\zsun$,
we obtain that the fraction  of NS-BH binary is {$(0.2$--$2.03)\,\times\, 10^{-4}$}.

The star formation rate of our galaxy at present 
is $1.65\,\msun$/yr~\citep{Licquia2015}.
Here, we use the generally accepted Salpeter-like initial mass function of the form 
\begin{equation}
  \phi (m)=C m^{-2.3} \,,
\end{equation}
which is appropriate for $m > 0.5\,\msun$~\citep{2019NatAs...3..482K}. 
Since the total mass of the new born stars in $5\times 10^7$\,yrs
is 
$8.3\times10^7\,\msun$,
we have the equality
\begin{equation}
    \int_{0.5\msun}^\infty m\phi (m) dm = 
     8.3\times10^7\,\msun
    \,.
\label{eq:total_mass}
\end{equation}
Then, the constant $C$
is determined as
\begin{equation}
     C=1.9\times 10^7\,\msun^{1.3} \,.
\end{equation} 
In the calculation of Paper I, we have considered only stars with mass larger than $5\,\msun$
because we are interested in NS-BHs.
Then, the total number of stars with mass larger than $5\,\msun$ is given by
\begin{equation}
    \int_{5\msun}^\infty \phi (m) dm=
    2.0\times10^6 \,.
\end{equation}
Assuming here that 50\% of the stellar systems are binaries,
the number of binaries is $6.7\times10^5$.
{Under this assumption, 
the number of the Pop I binary stars which evolve 
to PSR-BHs become $(0.2$--$2.03)\times 10^{-4} \times 6.7\times 10^5=13.4$--$136$ (see Table~\ref{T.PSR-BH}).}
\begin{table*}
\caption{{The numbers of galactic PSR-BHs for each model.
The total numbers of galactic PSR-BHs
and the numbers of observable PSR-BHs by future SKA observation for each model
are shown out of and in parentheses, respectively.}}
\label{T.PSR-BH}
\begin{center}
\begin{tabular}{ccccccccc}
\hline
  & Standard & \,$\sigma_k=500$\,km/s\, & \,$\alpha\lambda=0.1$\, &\,$\alpha\lambda=0.5$\, & \,$\alpha\lambda=10$\, & \,$\beta=0.1$\, & \,$\beta=0.5$\,& BH kick \\
 \hline
  PSR-BHs&   98.5 (19.7) &  21.4 (4.29)& 85.8 (17.2) & 136 (27.2)& 13.4 (2.68) & 40.2 (8.04) & 43.6 (8.71) & 29.5 (5.90)\\
 \hline
\end{tabular}
\end{center}
\end{table*}

Now PSRs are beaming so that only the fraction of
$\sim 0.2$ can be observed~\citep{2019ApJ...870...71P}
(see also~\cite{2015MNRAS.448..928K}).
While at present all the PSRs in our galaxy are not observed
but only $10\%$~\citep{2008LRR....11....8L} or so
(see also a recent review~\cite{2021arXiv210703915C})
due to the sensitivity of radio telescopes at present, 
in the SKA era almost all the active radio PSRs in our galaxy can be observed~\citep{2015aska.confE..40K}. 
{Thus, the expected number of active NS-BHs
in our galaxy found in radio band at present is {$0.268$--$2.72$.}}
This expected number for NS-BHs is marginal for detection 
which depends on the direction of magnetic field of the pulsar, the radio power, the distance to the pulsar and so on. 
Thus, the current no observation of NS-BH radio PSRs 
is more or less consistent. 

In relation to this marginal situation, it would be interesting to
point out the observation history of PSR J1740-3052 which was found in 2001~\citep{2001MNRAS.325..979S}. This binary PSR has a PSR period of 0.57\,sec and the characteristic age of $3.5\times 10^5$\,yrs with the orbital period of 231 days and the mass function of $8.7\,\msun$.
Thus the companion mass exceeds $11\,\msun$
so that it was not possible to rule out a BH companion in 2001 when this binary was found. At this time
PSR J1740-3052 could be a NS-BH binary formed in our above scenario for example. After 11 years from the discovery of PSR J1740-3052, \cite{2012MNRAS.425.2378M} found 
the main sequence star companion,
and concluded that PSR J1740-3052 is not a NS-BH binary. However, because the expected number of observable NS-BH binaries
in future SKA era is 
{2.68--27.2},
the existence of NS-BH binaries in our galaxy can be confirmed in future.

\begin{figure*}
\begin{tabular}{cc}
      \begin{minipage}[t]{0.43\hsize}
        \centering
\includegraphics[width=\hsize]{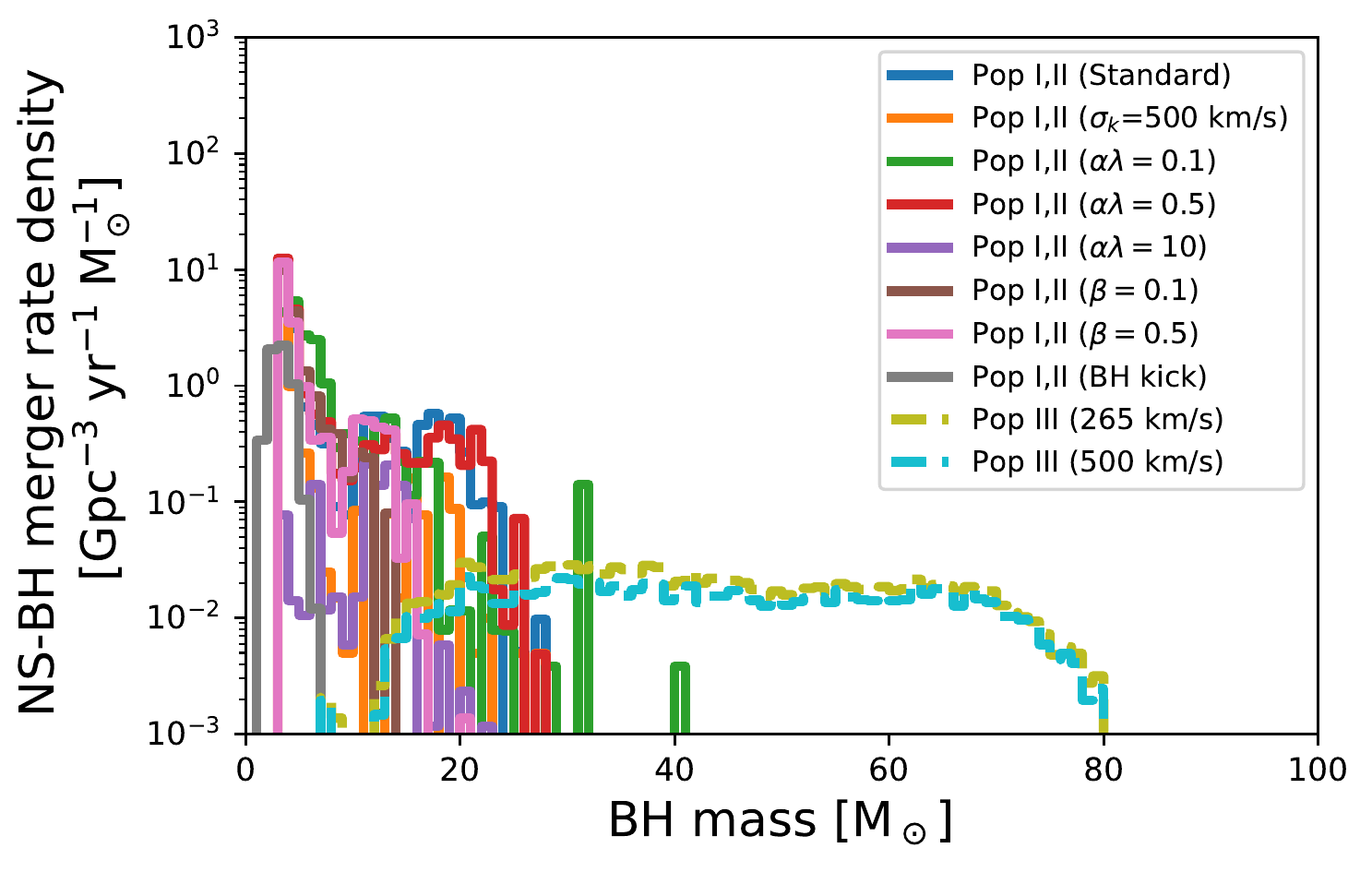}
\end{minipage} 
      \begin{minipage}[t]{0.43\hsize}
        \centering
\includegraphics[width=\hsize]{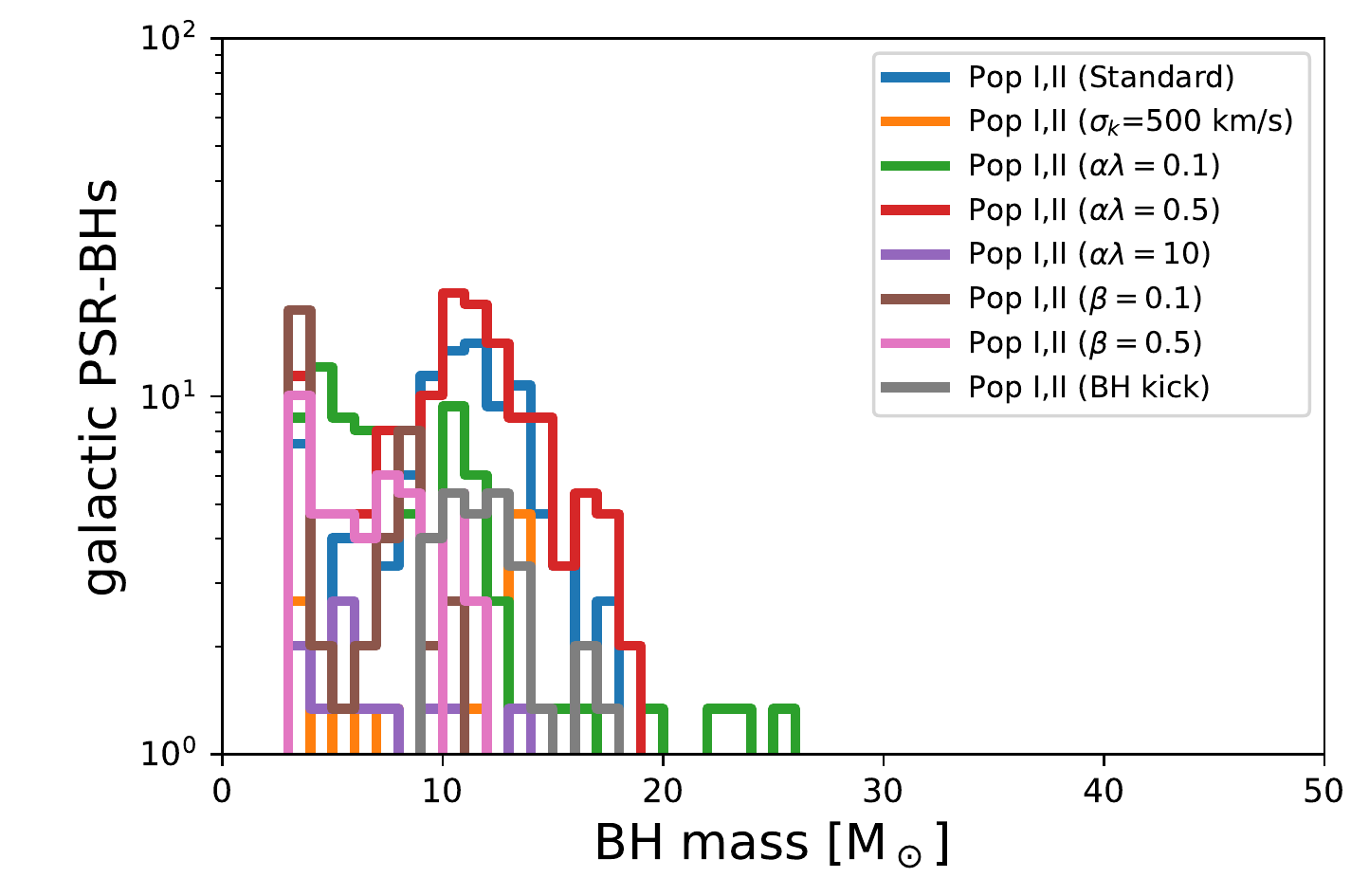}
\end{minipage} \\
    \end{tabular}
    \caption{{The left panel shows BH mass distributions of detectable NS-BHs at $z=0$. The sold lines are the cases of Pop I,II NS-BHs for each model.
    The dased lines are the case of Pop III NS-BHs.
    The right panel shows the BH mass distributions of galactic PSR-BHs for each model. }}
\label{fig:BH mass}
\end{figure*}

{Figure~\ref{fig:BH mass} shows the BH mass distributions of detectable NS-BH mergers at $z=0$ (the left panel)
and galactic PSR-BHs (the right panel). 
Since the galactic PSR-BHs are made by Pop I ($Z=\zsun$) stars, the BH masses of PSR-BHs are typically lower than those of NS-BHs which are detectable by the GWs.}  

\begin{figure}

\includegraphics[width=\hsize]{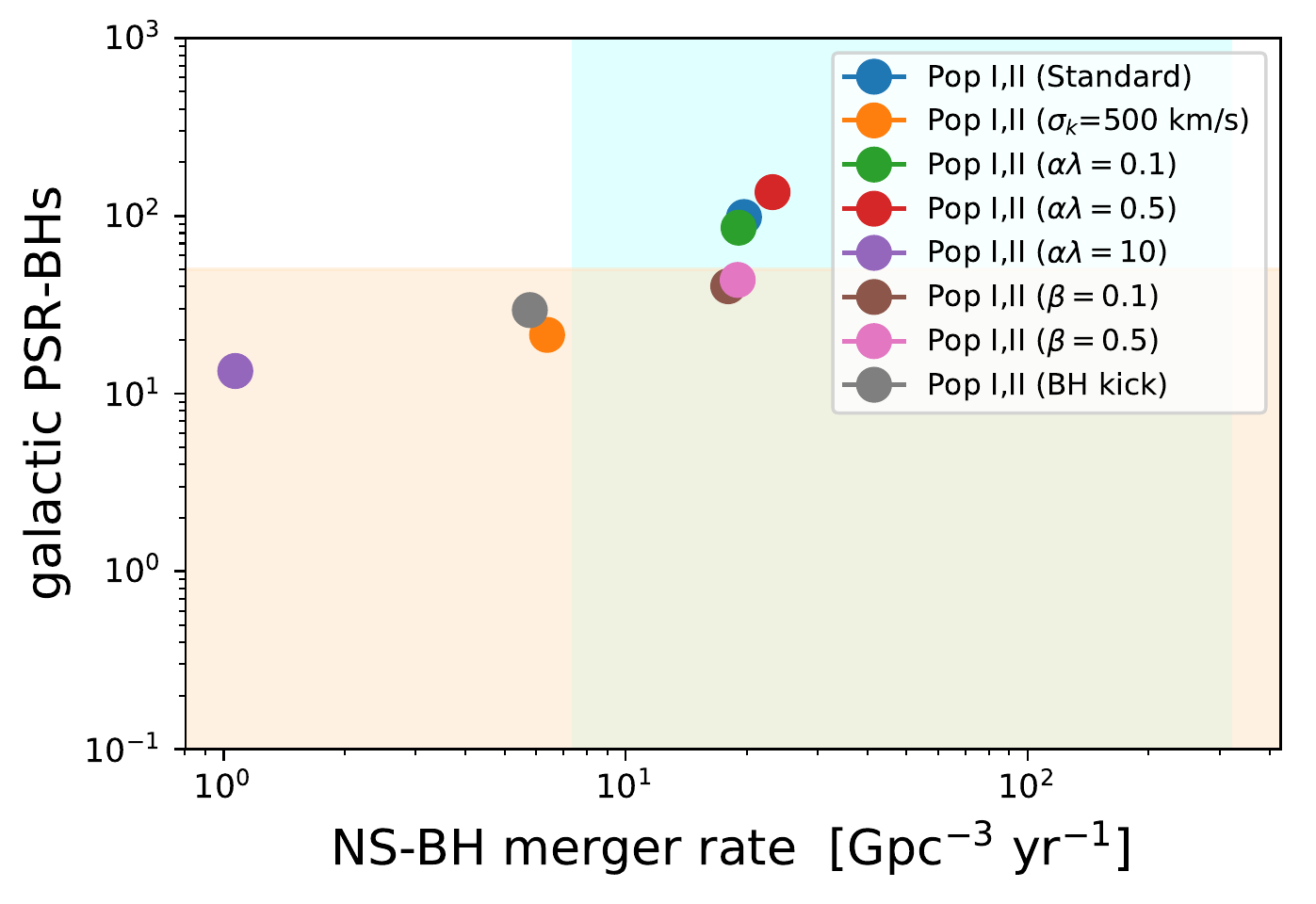}
    \caption{{The merger rates of NS-BHs and the numbers of galactic PSR-BHs for each model. 
    The blue shaded region shows the constraint of the NS-BH merger rate 7.4--320 $\rm yr^{-1} Gpc^{-3}$.
    The orange shaded region shows the constraint where the number of detectable galactic PSR-BHs by the present radio observation is less than 1.}}
\label{fig:rate}
\end{figure}

Figure~\ref{fig:rate} shows the merger rates of NS-BHs and the numbers of galactic PSR-BHs for each model. 
    The blue shaded region shows the constraint of the NS-BH merger rate of LIGO-Virgo result of 7.4--320 $\rm yr^{-1} Gpc^{-3}$.
    The orange shaded region shows the constraint where the number of detectable galactic PSR-BHs by the present radio observation is less than 1.
    Here, we define the upper limit of the number of detectable galactic PSR-BHs by the present radio observation is 1 because there is no observation of PSR-BH now. When the above upper limit becomes reality, 
    we can say the followings.
    {The $\beta=0.1$ model and the $\beta=0.5$ model satisfy these constraints.
    The standard model, the $\alpha\lambda=0.1$ model, and the $\alpha\lambda=0.5$ model are consistent with the NS-BH merger rate, but not consistent with the number of detectable PSR-BHs by the present radio observation slightly. On the other hand, the $\sigma_k=500$ km/s model and the BH kick model are consistent with the number of detectable PSR-BHs by the present radio observation, but not consistent with the NS-BH merger rate slightly. In the case of the $\alpha\lambda=10$ model, although it satisfies the constraint of the number of detectable PSR-BHs, the NS-BH merger rate of this model is about 10 times less than the lower limit of the LIGO result.}
    
     {At present the blue and orange shaded regions in Figure~\ref{fig:rate} are very large because of small number of NS-BHs observed by GW and none of PSR-BH observed by radio. However, they will increase so that we can get severe constraints on the possible theoretical model in future.}
     


\section{Discussion}

Some of NS-BHs which have a radio emission can be observed as PSR-BHs.
Our calculation shows that the SKA will detect 2.68--19.7 PSR-BHs in our galaxy.
We can check the consistency of the field binary model not only with the merger rate of NS-BHs, but also with the PSR-BH observations by future SKA.
{This can give a stronger constraint than Figure~\ref{fig:rate}.}
Since the SKA will find only young NS-BHs because the PSR lifetime is only $\sim 5\times 10^7$\,yrs, the origin of such NS-BHs will be confirmed as Pop I stars.
The maximum BH mass of Pop I PSR-BHs is $\lesssim20\,\msun$ (see Figure~\ref{fig:BH mass}). 
On the other hand, NS-BHs which are detected by GW observations are possibly the summation of each metallicity of Pop I, Pop II and Pop III.
The maximum BH mass of these NS-BHs is much more than $20\,\msun$ (see Figure~\ref{fig:BH mass}).
Thus, the comparison of the BH mass distribution of NS-BH binaries detected by GW observations with the BH mass distribution of PSR-BHs detected by the SKA might show the dependence of BH mass on the metallicity.

In our calculation, we have ignored the life prolongation of the spin due to accretion from a companion star like millisecond PSRs.
Here, $\sim 90\%$ of NS-BHs are formed as {\BHNS} where
{\BHNS} means that the primary star which is initially more massive than the companion star evolves to the BH first, and the secondary star evolves to the NS next.
In this case, accretion onto the NS does not occur.
On the other hand, $\sim 10\%$ of NS-BHs are formed as {\NSBH} which means that the primary star evolves to the NS first, and the secondary star evolves to the BH next.
In this case, progenitors of {\NSBH} evolve via HMXBs so that accretion onto the NS can occur. 
However the spin up due to accretion onto the NSs in HMXBs does not occur in general.
The magnetic field strength of the NSs in HMXBs is typically so large ($\sim 10^{12}$\,G)~\citep{Enoto2019} that the  matter cannot accrete directly onto the NS surface in the disc plane, but rather forms a funnel flow onto the magnetic poles because it is constrained to follow the magnetic field lines \citep[e.g.][]{Pringle1972}.
Thus, the PSR lifetime of {\NSBH} has been treated similar to  that of {\BHNS} in this Letter.

\section*{Acknowledgment}

We thank the anonymous referee and Wataru Ishizaki for useful comments.
T. K. acknowledges support from the University of Tokyo Young Excellent Researcher program and from JSPS KAKENHI Grant Number JP21K13915.
H. N. acknowledges support from JSPS KAKENHI Grant Numbers JP21K03582, JP21H01082 and JP17H06358.

\section*{Data Availability}

Results will be shared on reasonable request to corresponding author.

\bibliographystyle{mnras}

\bibliography{ref}

\end{document}